\documentclass[twocolumn,showpacs,preprintnumbers,amsmath,amssymb]{revtex4}
\usepackage{graphicx}% Include figure files
\usepackage{dcolumn}% Align table columns on decimal point
\usepackage{bm}% bold math
\usepackage{latexsym}
\usepackage{amsfonts}
\usepackage{amssymb}
\usepackage{amsmath}
\begin{document}

%\preprint{}

\title{Kaluza-Klein brane cosmology with a bulk scalar field}
\author{Arianto$^{(1,3)}$}
\email{arianto@fisika.unud.ac.id}
\author{F.P. Zen$^{(2,3)}$}
\email{fpzen@fi.itb.ac.id}
\author{S. Feranie$^{(4)}$}
\email{feranie@upi.edu}
\author{I P. Widyatmika$^{(1)}$}
\email{widyatmika@fisika.unud.ac.id}
\author{B.E. Gunara$^{(2,3)}$}
\email{bobby@fi.itb.ac.id}
\affiliation{$^{(1)}$Department of Physics, Faculty of Mathematics and Natural Sciences\\Udayana University\\
Jalan Kampus Bukit Jimbaran Kuta-Bali 80361, Indonesia.\\
$^{(2)}$Theoretical Physics Lab., THEPI Devision, and \\
$^{(3)}$Indonesia Center for Theoretical and Mathematical Physics (ICTMP)\\
Faculty of Mathematics and Natural Sciences,\\
 Institut Teknologi Bandung,\\
Jalan Ganesha 10 Bandung 40132, Indonesia. \\
$^{(4)}$Jurusan Pendidikan Fisika, FPMIPA,\\
Universitas Pendidikan Indonesia\\
Jalan Dr. Setiabudi 229, Bandung 40154, Indonesia. }

%===============================================================%
%************************* ABSTRACT ****************************%
%===============================================================%
\begin{abstract}
The brane-world cosmological model in higher-dimensional spacetime
with a bulk scalar field is investigated. We derive the
$(4+n)$-dimensional gravitational field equations for the scalar
field on the $(3+n)$-brane in a $(5+n)$-dimensional bulk with
Einstein gravity plus a self-interacting scalar field. The
$(4+n)$-dimensional gravitational field equations can be
formulated to standard form with the extra component. Using this
formalism we study the Kaluza-Klein brane cosmology. We derive the
Friedmann equation and a possible energy leak out of the brane
into the bulk. We present some exact solutions corresponding to
vacuum brane and matter on the brane.
\end{abstract}

\maketitle

%===============================================================%
%************************ SECTION I ****************************%
%===============================================================%
\section{Introduction}
The idea that our universe has dimensions more than four has been
around since the first attempts to unify fundamental forces. The
Kaluza-Klein theories were one of the first attempts towards this
direction. According to the Kaluza-Klein picture, extra dimensions
are compactified to a very small length scale (naturally the
Planck scale), and as a result spacetime appears to be effectively
four-dimensions, insofar as low energies are concerned. On the
other hand, inspired by the discovery of D-branes within string
theories, the brane-world scenario is now one of the most
important ideas in high energy physics. According to brane-world
scenario, our physical Universe is envisioned as a
four-dimensional hypersurface in a five-dimensional bulk
spacetime. The standard model matter is confined to the brane but
gravity, by its universal character, can propagate in the extra
dimensions.  Much efforts to reveal cosmology on the brane have
been done in the context of five-dimensional spacetime, especially
after the stimulating proposals by Randall and Sundrum (RS)
\cite{Randall:1999ee,Randall:1999vf}. In this model, a
five-dimensional realization of the Horava-Witten solution
\cite{Horava:1995qa}, the hierarchy problem can be solved by
introducing an appropriated exponential warp factor in the metric.
The various properties and characteristics of the RS model have
been extensively analyzed: the cosmology framework
\cite{Garriga:1999yh,Shiromizu:1999wj,
Langlois:2002bb,Brax:2004xh,Maartens:2010ar}, the low energy
effective theory
\cite{Mukohyama:2001ks,Wiseman:2002nn,Langlois:2002hz,Kanno:2002ia,Kanno:2002iaa,Shiromizu:2002qr,Brax:2003vf,Palma:2004fh,
Kobayashi:2006jw,CottaRamusino:2006iu,Fujii:2007fi,Arroja:2007ss},
the black hole physics
\cite{Chamblin:1999by,Emparan:1999wa,Chamblin:2000ra,Shiromizu:2000pg,Tamaki:2003bq,Kudoh:2003xz},
and the Lorentz violation
\cite{Csaki:2000dm,Stoica:2001qe,Libanov:2005yf,Libanov:2005nv,
Bertolami:2006bf,Ahmadi:2006cr,Ahmadi:2007tr,Koroteev:2009xd,Farakos:2009ka,Farakos:2009ui,Arianto:2009wc}.

The brane-world models with scalar field in the bulk has been
discussed by various authors, see~\cite{Maeda:2000wr} and
references therein. It is believed that in the unified theory
approach, a dilatonic gravitational scalar field term is required
in the Einstein-Hilbert action~\cite{Maeda:2000wr} . One of the
first motivations to introduce a bulk scalar field is to stabilize
the distance between the two branes~\cite{Kanti:2000rd} in the
context of the first model introduced by Randall and Sundrum. A
second, motivation for studying scalar fields in the bulk is due
the possibility that such a setup could provide some clue to solve
the famous cosmological constant problem. Models with inflation
driven by bulk scalar field have been studied and it is shown that
inflation is possible without inflaton on the
brane~\cite{Himemoto:2000nd}. Later the quantum fluctuations of
brane-inflation and reheating issues are also addressed in the
braneworld scenario with bulk scalar field~\cite{Yokoyama:2001nw}.
The creation of a brane world with a bulk scalar field using an
instanton solution in a five dimensional Euclidean Einstein
equations is also considered~\cite{Aoyanagi:2006wd}.

In this paper, our main purpose is to construct brane cosmological
models in higher-dimensional spacetime with a bulk scalar field.
We generalize the case four-dimensional brane models to
$(4+n)$-dimensional brane models where $n$ represents internal
dimensions of the brane. In order to obtain four-dimensional
description of our Universe, we combine two representative and
contrastive approachs, that is, the Randall-Sundrum and the
Kaluza-Klein scenario. In the former case, the compactification is
done by the localization of the configuration along the extra
dimension, while in the latter one the reduction is achieved by
the compactification of the internal space of the brane. Such a
way of construction is called Kaluza-Klein brane-world, and it has
been investigated by various authors
\cite{Kanti:2001vb,Charmousis:2004zd,CuadrosMelgar:2005ex,Papantonopoulos:2006uj,
Chatillon:2006vw,Yamauchi:2007wm,Kanno:2007wj,Minamitsuji:2008mn,Feranie:2010th}.

This paper is organized as follows. In Section~\ref{sec:Effective
Equations}, we study a higher brane-world model in a $(5 +
n)$-dimensional spacetime with a bulk scalar field. We derive the
$(4 + n)$-dimensional Einstein equations using the geometrical
approach. We transform the non-conventional kinetic term in the
Einstein equations into the standard form. In Section
\ref{sec:Braneworld cosmology}, the Kaluza-Klein brane worlds
cosmology is presented by using the extra component formalism.
Then we discuss the vacuum brane solutions with an initial
singularity and a possible energy leak out of the brane into the
bulk. For completeness, we study the static internal dimension
with matter on the brane in Section~\ref{sec:staticint}. We derive
the Friedmann equation and a possible energy leak out of the brane
into the bulk. In section \ref{sec:E-frame}, we study dimensional reduction of the
higher-dimensional theory and then performing a conformal transformation in
order to recover ordinary 4-dimensional general relativity.
Section \ref{sec:conclusion} is devoted to the conclusions.

%===============================================================%
%************************ SECTION II ***************************%
%===============================================================%
\section{\label{sec:Effective Equations}Effective gravitational equations}

\subsection{Action and field equations}
We consider the higher-dimensional dilatonic brane-world, i.e. the
higher-dimensional brane-world where the bulk is allowed to
contain a single non-gravitational degree of freedom, the scalar
field $\phi$. A ($3+n$)-brane with the ($4+n$)-dimensional
spacetime embedded in the ($5+n$)-dimensional spacetime is located
at $y=0$, where the $y$ direction is compactified on the orbifold.
This higher-dimensional dilatonic brane-world model is described
by the action
\begin{eqnarray}
    S &=&  \int d^{5+n} x \sqrt{-g} \left[\frac{1}{2\kappa^2}{\cal R}
     -{1\over 2}g^{ab}\nabla_a\phi\nabla_b\phi - V(\phi) \right]\nonumber\\
     &&+  \int d^{4+n}x \sqrt{-h}\left[-\sigma(\phi) + {\cal L}_{m}\right],
     \label{eq:action}
\end{eqnarray}
where $\mathcal{R}$ is the Ricci scalar of the $(5+n)$-dimensional
metric $g_{ab}$. $\sigma$ is the brane tension which is allowed to
the function of $\phi$, and ${\cal L}_{m}$ is the Lagrangian for
matter localized on the brane. A metric $h_{\mu\nu}$ is the
induced metric on the brane.

We write the coordinate system for the bulk spacetime in the form
\begin{equation}
    ds^2= g_{ab} dx^a dx^b = dy^2 +  g_{\mu\nu}(y,x)dx^\mu dx^\nu,
    \label{eq:metric}
\end{equation}
and we may assume that the position of the brane is $y=0$ in this
coordinate system so that the induced metric on the brane is
$h_{\mu\nu}(x)=g_{\mu\nu}(y=0,x)$ and the extrinsic curvature is
defined as
\begin{equation}
    K_{\mu\nu} =-{1\over 2}\frac{\partial}{\partial y}g_{\mu\nu} \equiv -{1\over 2}g_{\mu\nu,y}.
    \label{eq:ext-curv}
\end{equation}

The Einstein equations can be obtained by varying the action
(\ref{eq:action}) with respect to the gravitational field
$g_{ab}$,
\begin{eqnarray}
    &&{\cal G}_{ab} = \kappa^2 {\cal T}_{ab} + \kappa^2 \delta_a^\mu \delta_b^\nu (-\sigma g_{\mu\nu}+t_{\mu\nu}) \delta(y),
   \label{eq:einstein-eq}
\end{eqnarray}
where ${\cal G}_{ab}={\cal R}_{ab}-g_{ab}{\cal R}/2$ is the
($5+n$)-dimensional Einstein tensor, and ${\cal T}_{ab}$ is the
energy--momentum tensor for the scalar field,
\begin{equation}
    {\cal T}_{ab} = \nabla_a\phi\nabla_b\phi - g_{ab}\left({1\over 2}g^{cd}\nabla_c\phi\nabla_d\phi + V(\phi)\right),
\end{equation}
while the total energy--momentum tensor on the brane $t_{\mu\nu}$
is defined as
\begin{equation}
    t_{\mu\nu} \equiv -2{\delta {\cal L}_m \over \delta g^{\mu\nu}} +  g_{\mu\nu} {\cal L}_m.
\end{equation}
The equation of motion for the scalar field reads:
\begin{equation}
    \nabla^a\nabla_a\phi-V'(\phi)-\sigma'(\phi)\delta(y) =0,
   \label{eq:eos-scalar}
\end{equation}
where a prime denotes a derivative with respect to $\phi$.

In the coordinate system (\ref{eq:metric}) and by the application
of the Gauss-Codazzi equations, one has the ($5+n$)-dimensional
field equations as follows
\begin{eqnarray}
    {\cal G}^y_{\ y} &=& -{1\over 2}R + {1\over 2}K^2 - {1\over 2}K^{\alpha\beta}K_{\alpha\beta} = \kappa^2 {\cal T}^y_{\ y},
    \label{eq:einstein-eq-yycomp}\\
   {\cal G}^y_{\ \mu} &=& - \nabla_\alpha K_{\mu}{}^{\alpha}+\nabla_{\mu}K =\kappa^2 {\cal T}^y_{\ \mu},
   \label{eq:einstein-eq-ymucomp}\\
   {\cal G}^\mu_{\ \nu} &=& G^\mu_{\ \nu} + (K^\mu_{\ \nu}-\delta^\mu_{\ \nu}K)_{,y} - KK^\mu_{\ \nu} \nonumber\\
   &&+{1\over 2}\delta^\mu_{\ \nu}(K^2 + K^{\alpha\beta}K_{\alpha\beta})\nonumber\\
   &=& \kappa^2 {\cal T}^\mu_{\ \nu}
   + \kappa^2\left(-\sigma \delta^\mu_{\ \nu} + t^\mu_{\ \nu}\right)\delta(y),
   \label{eq:einstein-eq-munucomp}
\end{eqnarray}
where $G^\mu_{\ \nu }=R^\mu_{\ \nu }-\delta^\mu_{\ \nu }R/2$ is
the $(4+n)$-dimensional Einstein tensor and the covariant
derivative $\nabla_\mu$ is calculated with respect to the metric
$g_{\mu\nu}$. Combining equations (\ref{eq:einstein-eq-yycomp})
with (\ref{eq:einstein-eq-munucomp}), and using the
$(5+n)$-dimensional Weyl tensor, we obtain the following
$(4+n)$-dimensional Einstein equations
\begin{eqnarray}
    G^{\mu}_{\ \nu } &=&  - K^{\mu\alpha} K_{\alpha\nu}+ KK^{\mu}_{\ \nu }
    + {1\over 2}\delta^{\mu}_{\ \nu }\left(K_{\alpha\beta}K^{\alpha\beta}- K^2\right) \nonumber\\
    && + {2+n\over 3+n}\kappa^2\left[{\cal T}^{\mu}_{\ \nu } - {1 \over 4+n}\delta^{\mu}_{\ \nu } {\cal T}^{\alpha}_{\ \alpha}
    + {3+n\over 4+n}\delta^{\mu}_{\ \nu } {\cal T}^y_{\ y}\right]\nonumber\\
    &&- E^{\mu}_{\ \nu },
   \label{eq:einstein-proj}
\end{eqnarray}
Here, we have defined that the term ${\cal T}^{\alpha}{}_{\alpha}$
is the trace defined with respect to the $(4+n)$-dimensional
metric $g_{\mu\nu}$, and the projected Weyl tensor is defined as
$E_{\mu\nu}=C_{y\mu y\nu}|_{y=0}$. To eliminate the extrinsic
curvature, we need the junction conditions. It can be obtained by
collecting together the terms in field equations which contain a
$\delta$-function. By assuming $Z_2$-symmetry we find
\begin{eqnarray}
    \left[ K^\mu_{\ \nu} - \delta^{\mu}_{\ \nu} K \right] |_{y=0}&=&
    {\kappa^2 \over 2} \left(-\sigma \delta^{\mu}_{\ \nu} + t^\mu_{\ \nu}\right),
    \label{eq:JC-pr}\\
    \left[ \partial_y \phi \right] |_{y=0} &=&{1\over 2}\sigma'(\phi).
    \label{eq:JC-scalar}
\end{eqnarray}
Using this junction conditions, we find the $(4+n)$-dimensional
generalization of the Shiromizu-Maeda-Sasaki equations
\cite{Shiromizu:1999wj},
\begin{eqnarray}
    G^{\mu}_{\ \nu} &=& \frac{(2+n)\kappa^4}{4(3+n)}\sigma t^{\mu}_{\ \nu}\nonumber\\
    &&+ \frac{(2+n)\kappa^2}{(3+n)}\left[\nabla^\mu\phi \nabla_\nu\phi-{(5+n)\over 2(4+n)}\delta^{\mu}_{\ \nu}\nabla^\alpha\phi
    \nabla_\alpha\phi\right]\nonumber\\
    && - \Lambda_b\delta^{\mu}_{\ \nu} + \kappa^4\pi^{\mu}_{\ \nu} - E^{\mu}_{\ \nu},
   \label{eq:einstein-brane}
\end{eqnarray}
where the induced cosmological constant on the brane is given by
\begin{eqnarray}
    \Lambda_b(\phi) &=& \frac{(2+n)\kappa^2}{(4+n)}\left[V + \frac{(4+n)\kappa^2}{8(3+n)}\sigma^2 -
    \frac{1}{8}\sigma'^2\right],
    \label{eq:cos-brane}
\end{eqnarray}
and the local quadratic energy-momentum tensor on the brane is
\begin{eqnarray}
    \pi^{\mu}_{\ \nu} &=& - {1\over 4}t^{\mu}_{\ \alpha} t^{\alpha}_{\ \nu}
    + {1\over 4(3+n)}tt^{\mu}_{\ \nu}  \nonumber\\
    &&+ {1\over 8} \delta^{\mu}_{\ \nu} \left(t_{\alpha\beta}t^{\alpha\beta}-
    {1\over 3+n}t^2\right).
    \label{eq:qud-matter}
\end{eqnarray}
We note that Eq.~(\ref{eq:cos-brane}) may not be constant in
general, as is clear from its expression. The first term comes
from the scalar field potential. The second term is contribution
from the brane tension, which yields extrinsic curvature of the
brane and its quadratic. The third term is the first derivative of
the brane tension, which leads to a discontinuity in the scalar
field gradient normal to the brane.

The scalar field equation of motion is given by
\begin{equation}
   \nabla^\alpha\nabla_\alpha\phi - {1\over \kappa^2}\Lambda'_b(\phi) = J_n,
  \label{eq:eos}
\end{equation}
where a possible energy leak out of the brane into the bulk,
$J_n$, is defined by
\begin{eqnarray}
   J_n &=& \frac{(2+n)}{4(4+n)}\left(\sigma'\sigma'' - \frac{\kappa^2}{3+n}\sigma t\right)
   + \frac{2}{4+n}\nabla^\alpha\nabla_\alpha\phi \nonumber\\
    &&- \frac{2+n}{4+n} [\partial^2_y\phi]_{y=0}.
  \label{eq:j}
\end{eqnarray}

\subsection{Extra component formalism}
The induced Einstein equations on the brane,
Eq.~(\ref{eq:einstein-brane}), contain the non-conventional
kinetic term. However, we can transform the non-conventional
kinetic term into the standard form, then
Eq.~(\ref{eq:einstein-brane}) can be rewritten as
\begin{eqnarray}
    G^{\mu}_{\ \nu} &=& \kappa^2\left(\nabla^\mu\phi \nabla_\nu\phi-{1\over 2}\delta^{\mu}_{\ \nu}\nabla^\alpha\phi \nabla_\alpha\phi\right)
    -\kappa^2 V_{eff}\delta^{\mu}_{\ \nu} \nonumber\\
    &&+  X^{\mu}_{\ \nu},
   \label{eq:einstein-stdform}
\end{eqnarray}
where the extra component $X^{\mu}_{\ \nu}$ is defined as
\begin{equation}
    X^{\mu}_{\ \nu} \equiv Y^{\mu}_{\ \nu} - \cal{Z}^{\mu}_{\ \nu},
   \label{eq:x-matter}
\end{equation}
with
\begin{equation}
    Y^{\mu}_{\ \nu} \equiv \frac{(2+n)\kappa^4}{4(3+n)}\sigma t^{\mu}_{\ \nu} + \kappa^4\pi^{\mu}_{ \nu},
   \label{eq:y-matter}
\end{equation}
and
\begin{equation}
   {\cal{Z}}^{\mu}_{\ \nu} \equiv E^{\mu}_{\ \nu}
    + \frac{1}{(3+n)}\left(\nabla^\mu\phi \nabla_\nu\phi-{1\over 4+n}\delta^{\mu}_{\ \nu}\nabla^\alpha\phi
    \nabla_\alpha\phi\right).
   \label{eq:z-matter}
\end{equation}

Following Eq.~(\ref{eq:eos}) we have defined the effective
potential $V_{eff}=\Lambda_b/\kappa^2$, where the induced
cosmological constant on the brane $\Lambda_b$ is given by
Eq.~(\ref{eq:cos-brane}). Then the scalar field equation of motion
becomes
\begin{equation}
   \nabla^\alpha\nabla_\alpha\phi - V'_{eff} = J_n.
  \label{eq:eos-std}
\end{equation}
From Eq.~(\ref{eq:einstein-stdform}), the Bianchi identity implies
\begin{equation}
    \nabla_\mu X^{\mu}_{\ \nu} = -\kappa^2 J_n \nabla_\nu\phi.
   \label{eq:concv-matter}
\end{equation}
Note that the energy-momentum tensor of the extra component is not
conserved due to the existence of the bulk scalar field. In the
absence of matter on the brane, $Y^{\mu}_{\ \nu}=0$, then we have
$X^{\mu}_{\ \mu}=-{\cal Z}^{\mu}_{\ \mu}=0$. In this case, we can
interpret the extra component as the energy-momentum tensor for
the dark radiation.

In the following section, we attempt to study analytically the
cosmological consequences of higher-dimensional brane-world. We
used the extra component formalism to discuss cosmology in the
brane Universe in the context of the Kaluza-Klein brane-world
scheme, i.e., to consider Kaluza-Klein compactifications on the
brane.

%%%%%%%%%%%%%%%%%%%%%%%%%%%%%%%%%%%%%%%%%%%%%%%%%%%%%%%%%%%%%%%%%%
%%%%%%%%%%%%%%%%%%%%%% SECTION III %%%%%%%%%%%%%%%%%%%%%%%%%%%%%%%
\section{\label{sec:Braneworld cosmology}Dilatonic Kaluza-Klein Brane-world cosmology}

For the cosmological applications, we are interested in
homogeneous and isotropic geometries on the brane, hence the
metric on the brane is taken as the Friedmann-Robertson-Walker
metric,
\begin{eqnarray}
    ds^2 = -dt^2 + a^2(t)\delta_{ij} dx^i dx^j + b^2(t)\delta_{\alpha\beta} dz^\alpha dz^\beta,
    \label{eq:metric-cos}
\end{eqnarray}
where $\delta_{ij}$ represents the metric of three-dimensional
ordinary spaces with the spatial coordinates $x^i$ ($i=1,2,3$),
while $\delta_{\alpha\beta}$ represents the metric of
$n$-dimensional compact spaces with the coordinates $z^\alpha$
($\alpha = 4, \ldots, 3+n$). We assume that the internal space is
given by $n$-dimensional torus. The scale factor $b$ denotes the
size of the internal dimensions, while the scale factor $a$ is the
usual scale factor for the external space.

In the background metric (\ref{eq:metric-cos}) and assuming that
$\phi$ only depends on time, we obtain the following equations
\begin{eqnarray}
   && G^0_{\ 0} =  -\kappa^2\left({1\over 2}\dot{\phi}^2 + V_{eff}\right) +  X^0_{\ 0},
   \label{eq:einstein-00}\\
   && G^i_{\ j} = \kappa^2\left({1\over 2}\dot{\phi}^2 - V_{eff}\right)\delta^i_{\ j} +  X^i_{\ j},
   \label{eq:einstein-ij}\\
   && G^\alpha_{\ \beta} = \kappa^2\left({1\over 2}\dot{\phi}^2 - V_{eff}\right)\delta^\alpha_{\ \beta} + X^\alpha_{\ \beta},
   \label{eq:einstein-ab}\\
   &&\ddot{\phi} + 3H_a\dot{\phi} + nH_b\dot{\phi} + V'_{eff} = -J_n,
    \label{eq:cos-eos}\\
   && \dot{t}^0_{\ 0} + 3H_a \left(t^0_{\ 0} -t^1_{\ 1} \right)  + nH_b\left(t^0_{\ 0} -t^4_{\ 4} \right) =0,
    \label{eq:conservation}
\end{eqnarray}
where $H_a=\dot{a}/a$ and $H_b=\dot{b}/b$, and the components of
the $(4+n)$-dimensional Einstein tensor are
\begin{widetext}
\begin{eqnarray}
    G^0_{\ 0} &=& -3\left(H^2_a +{k_a\over a^2}\right) -n\left[{(n-1)\over 2}H^2_b + 3H_aH_b+ {(n-1)\over 2}{k_b\over b^2}\right],
    \label{eq:einstein-tens-00}\\
    G^i_{\ j} &=& -\left( 2\dot{H}_a + 3H^2_a +{k_a\over a^2}\right)\delta^i_{\ j}
    - n\left[\dot{H}_b +{(n+1)\over 2}H^2_b +2H_a H_b + {(n-1)\over 2}{k_b\over b^2} \right]\delta^i_{\ j}\ ,
   \label{eq:einstein-tens-ij}\\
    G^\alpha_{\ \beta} &=& -3\left(\dot{H}_a + 2H^2_a +{k_a\over a^2}\right)\delta^\alpha_{\ \beta}
    - (n-1)\left[ \dot{H}_b + {n\over 2}H^2_b + 3H_a H_b + {(n-2)\over 2}{k_b\over b^2} \right]\delta^\alpha_{\ \beta}\ .
   \label{eq:einstein-tens-ab}
\end{eqnarray}
\end{widetext}
The values of $k_a$ and $k_b$ are related to the curvatures of the
external space and the internal space, respectively. In this paper
we assume $k_b = 0$ for simplicity. Equations
(\ref{eq:einstein-00})-(\ref{eq:conservation}) are our basic
equations to study cosmology in the Kaluza-Klein brane-world.

The constraint equation for the extra component is given by
\begin{equation}
    \dot{X}^0_{\ 0} + 3H_a \left(X^0_{\ 0} -X^1_{\ 1} \right)  + nH_b\left(X^0_{\ 0} -X^4_{\ 4} \right)
    =-\kappa^2 J_n \dot{\phi}.
    \label{eq:X-constraint}
\end{equation}
By combining Eqs.~(\ref{eq:einstein-00}), (\ref{eq:einstein-ij}),
and (\ref{eq:einstein-ab}) one can write
\begin{equation}
    X^0_{\ 0} + 3 X^1_{\ 1} - 2 X^4_{\ 4} = G^0_{\ 0} + 3 G^1_{\ 1} - 2 G^4_{\ 4} +
    2\kappa^2 V_{eff}.
    \label{eq:X-cons-1}
\end{equation}
%
%%%%%%%%%%%%%%%%%%%%%%%%%%%%%%%%%%%%%%%%%%%%%%%%%%%%%%%%%%%%%%%%%%
%%%%%%%%%%%%%%%%%%%%%% SUBSECTION %%%%%%%%%%%%%%%%%%%%%%%%%%%%%%%%
\subsection{Solutions with $\Lambda_b=0$ and $J_n=0$}

Let us first consider that the bulk and brane potentials obey the
generalized Randall-Sundrum condition $\Lambda_b=0$, so that
$V_{eff}=0$, and the energy conservation for scalar field on the
brane is satisfied, $J_n=0$. The solution of
Eq.~(\ref{eq:cos-eos}) is given by
\begin{eqnarray}
    \dot{\phi} = \frac{c_{\phi}}{a^3 b^n},
    \label{eq:cos-eos-sol}
\end{eqnarray}
where $c_{\phi}$ is an integration constant. Note that the scalar
field depends on both the external scale factor and the internal
scale factor, $\phi(t)=\phi(a(t),b(t))$.

In the following we consider a special case in which the
brane-world evolves with two scale factors. We take a simple
relation between the scale factors on the brane of the form
$b(t)=a^{\gamma}(t)$, where $\gamma$ is a constant. For the
internal scale factor $b(t)$ to be small compared to the external
scale factor $a(t)$, the constant $\gamma$ should be negative. In
this choice we have $H_b=\gamma H_a\equiv\gamma H$, and then
Eq.~(\ref{eq:X-constraint}) becomes
\begin{equation}
    \dot{X}^0_{\ 0} + \left(4 + n\gamma\right)H X^0_{\ 0}
    =-n\gamma(1-\gamma)\left[ \dot{H} +(3 +n\gamma)H^2\right]H.
    \label{eq:X-constraint-2}
\end{equation}
The Friedmann equation is given by
\begin{equation}
    3(1+\alpha_0)H^2 =  -3{k_a\over a^2}
    +{\kappa^2\over 2}\frac{c^2_{\phi}}{a^{2(3+n\gamma)}}  -  X^0_{\ 0},
    \label{eq:friedmann-vac}
\end{equation}
where
\begin{equation}
    \alpha_0 =  \frac{n\gamma(6-\gamma+n\gamma)}{6}.
    \label{eq:a}
\end{equation}
The extra component $X^0_{\ 0}$ is determined by the solution of
the equation (\ref{eq:X-constraint-2}) which can be rewritten as
\begin{eqnarray}
    \dot{X}^0_{\ 0} + \left(4 + \beta_0\right)X^0_{\ 0}H = \beta_1\frac{k_a}{a^2}H,
    \label{eq:X-constraint-3}
\end{eqnarray}
where
\begin{eqnarray}
    &&\beta_0=\frac{n\gamma(2+\gamma+n\gamma)}{3+n\gamma},\quad
    \beta_1=\frac{6n\gamma(1-\gamma)}{3+n\gamma},\nonumber\\
    &&\beta_2=\frac{3n(1-\gamma)(n-1)(4+\gamma+3n\gamma)}{4(3+n\gamma)}.
    \label{eq:b}
\end{eqnarray}
Note that the effect of internal dimension is given by definitions
(\ref{eq:a}) and (\ref{eq:b}). Equation (\ref{eq:X-constraint-3})
can be integrated to gives
\begin{equation}
    X^0_{\ 0} = \frac{\beta_1}{2+\beta_0}\frac{k_a}{a^2}-\frac{\varepsilon_0}{a^{4+\beta_0}},
    \label{eq:X-constraint-3-sol}
\end{equation}
where $\varepsilon_0$ is an integration constant. Inserting
Eq.~(\ref{eq:X-constraint-3-sol}) into
Eq.~(\ref{eq:friedmann-vac}), we find the modified Friedmann
equation
\begin{eqnarray}
    3(1+\alpha_0)H^2 &=&  -3\left[1+ \frac{\beta_1}{3(2+\beta_0)}\right]\frac{k_a}{a^2}\nonumber\\
    &&+{\kappa^2\over 2}\frac{c^2_{\phi}}{a^{2(3+n\gamma)}}
    +\frac{\varepsilon_0}{a^{4+\beta_0}}.
    \label{eq:friedmann-vac-1}
\end{eqnarray}

In what follows we study the solutions with an initial singularity
($a=0$). Using the conformal time variable $\eta$ defined by the
differential relation $dt=ad\eta$, we can rewrite
Eq.~(\ref{eq:friedmann-vac-1}) as
\begin{eqnarray}
    3(1+\alpha_0)\left({da\over d\eta}\right)^2 &=&-3\left[1+ \frac{\beta_1}{3(2+\beta_0)}\right]k_{a} a^2\nonumber\\
    &&+{\kappa^2\over 2}\frac{c^2_{\phi}}{a^{2(1+n\gamma)}}
    +\frac{\varepsilon_0}{a^{\beta_0}},
    \label{eq:friedmann-vac-confm}
\end{eqnarray}
and Eq.~(\ref{eq:cos-eos-sol}) as
\begin{equation}
    {d\phi\over d\eta} =\frac{c_\phi}{a^{(2+n\gamma)}}.
    \label{eq:scalar-confm}
\end{equation}
In the following we consider three cases: $\gamma=0,\pm 1$.

In the case $\gamma=-1$, the internal scale factor $b(t)$ is
proportional to $b(t)=1/a(t)$. We have $a\rightarrow 0$, and the
infinitely large internal space, $b\rightarrow\infty$.

In the case $\gamma=1$, the internal scale factor $b(t)$ is
related to $a(t)$ as $b(t)=a(t)$. The external scale factor
evolves as
\begin{equation}
    a^{2+n}(\tau) =\frac{\varepsilon_0\tau\left(\tau+\tau_c\right)}{3(1+k_a\tau^2)},
    \label{eq:friedmann-vac-confm-2}
\end{equation}
where the new variable time is given by
\begin{equation}
    \tau(\eta) =\left\{%
\begin{array}{ll}
    \sqrt{{3(2+n)\over 2(3+n)}}~|\eta|, & \hbox{for}~~k_a = 0\\
    \tan \sqrt{{3(2+n)\over 2(3+n)}}~|\eta|, & \hbox{for}~~k_a = +1 \\
    \tanh \sqrt{{3(2+n)\over 2(3+n)}}~|\eta|, & \hbox{for}~~k_a = -1 \\
\end{array}%
\right.,
    \label{eq:var-time-2}
\end{equation}
and
\begin{equation}
    \tau_c = \frac{\sqrt{6}\kappa|c_\phi|}{\varepsilon_0}.
    \label{eq:crh}
\end{equation}
The scalar field evolves as
\begin{equation}
    \phi -\phi_0 =\pm\sqrt{3+n\over (2+n)\kappa^2}\ln\left|{\tau\over \varepsilon_0(\tau + \tau_c)}\right|,
    \label{eq:scalar-confm-2}
\end{equation}
for $\varepsilon_0\neq 0$, or
\begin{equation}
    \phi -\phi_0 =\pm \sqrt{3+n\over (2+n)\kappa^2}\ln\left|\tau\right|,
    \label{eq:scalar-confm-2}
\end{equation}
for $\varepsilon_0 = 0$, where $\pm$ corresponds to the sign of
$c_\phi$. Note that the initial singularity appears at $\tau = 0$.
If we convert to proper time the evolution of the scale factor
with $c_\phi\neq 0$ is given by
\begin{equation}
    a(t) = \left({2\over 3}\right)^{1\over 2(2+n)}\left({3+n\over 2+n}\right)^{5+n\over 2(2+n)(3+n)}
    \left(\kappa c_\phi\right)^{1\over 3+n}t^{1\over 3+n}.
    \label{eq:scale-proper}
\end{equation}
Then, the Universe starts expanding away from the initial
singularity.

More interesting case is in the static internal dimension,
$\gamma=0$, we obtain
\begin{equation}
    a^2(\tau) =\frac{\varepsilon_0\tau\left(\tau+\tau_c\right)}{3(1+k_a\tau^2)},
    \label{eq:friedmann-vac-confm-1}
\end{equation}
where
\begin{equation}
    \tau(\eta) =\left\{%
\begin{array}{ll}
    |\eta|, & \hbox{for}~~k_a = 0\\
    \tan |\eta|, & \hbox{for}~~k_a = +1 \\
    \tanh |\eta|, & \hbox{for}~~k_a = -1 \\
\end{array}%
\right.,
    \label{eq:var-time-1}
\end{equation}
and
\begin{equation}
    \phi -\phi_0 =\pm\sqrt{3\over 2\kappa^2}\ln\left|{\tau\over
    \varepsilon_0(\tau + \tau_c)}\right|,
    \label{eq:scalar-confm-3}
\end{equation}
for $\varepsilon_0\neq 0$, or
\begin{equation}
    \phi -\phi_0 =\pm\sqrt{3\over 2\kappa^2}\ln\left|\tau\right|,
    \label{eq:scalar-confm-3}
\end{equation}
for $\varepsilon_0 = 0$. In a spatially flat Universe the
evolution of the scale factor is the same as the standard
cosmological solution with stiff matter,
\begin{equation}
    a(t) = \left({3\over 2}c_\phi^2\kappa^2\right)^{1/6}t^{1/3}.
    \label{eq:scale-proper-1}
\end{equation}
We also find that the Universe starts expanding away from the
initial singularity.

%%%%%%%%%%%%%%%%%%%%%%%%%%%%%%%%%%%%%%%%%%%%%%%%%%%%%%%%%%%%%%%%%%
%%%%%%%%%%%%%%%%%%%%%% SUBSECTION %%%%%%%%%%%%%%%%%%%%%%%%%%%%%%%%
\subsection{Solutions with exponential potential}

We may now consider the case when the bulk scalar field potential
and the brane tension are given as exponents, with some constant
parameters $V_0$, $\sigma_0$ and $\lambda$,
\begin{eqnarray}
    V(\phi) &=& V_0\exp\left(-\frac{2}{3+n}\lambda\kappa\phi\right),
    \label{eq:pot}\\
    \sigma(\phi) &=& \sigma_0\exp\left(-\frac{1}{3+n}\lambda\kappa\phi\right).
    \label{eq:tension}
\end{eqnarray}
These are string-inspired values for the
potentials~\cite{Chamblin:1999ya,Langlois:2003dd,Koyama:2003yz}.
Then the effective potential is given by
\begin{equation}
    V_{eff} = V_{eff,0}\exp\left(-\frac{2}{3+n}\lambda\kappa\phi\right)={\Lambda_b\over \kappa^2},
    \label{eq:pot-eff}
\end{equation}
where
\begin{equation}
    V_{eff,0}=\frac{2+n}{4+n}V_0 +
    \frac{(2+n)\kappa^2\sigma_0^2}{8(3+n)}\left[1-\frac{\lambda^2}{3(4+n)}\right].
    \label{eq:pot-eff-0}
\end{equation}
We also assume the proportionality relation between the scalar
field and the logarithm of the scale factors,
\begin{equation}
    \phi=\frac{\lambda}{\kappa}\ln(ab)=\frac{\lambda}{\kappa}\ln(a^{1+\gamma}).
    \label{eq:pot-eff-0}
\end{equation}
From the above model, if $\lambda =0$ we have $\phi=0$ and then
$V=V_0$, $\sigma=\sigma_0$, and $V_{eff} = V_{eff,0}$. The induced
cosmological constant on the brane, $\Lambda_b$, is exactly
constant. If we set $\Lambda_b=0$, the generalized Randall-Sundrum
condition is given by
\begin{equation}
    \sigma_0=\sqrt{\frac{8(3+n)}{(4+n)\kappa}(-V_0)}.
    \label{eq:fine-tune}
\end{equation}
For the positive brane tension, $V_0$ must be negative. Thus $V_0$
can be interpreted as a bulk cosmological constant and the
Randall-Sundrum brane is a slice in (5+n)-dimensional anti-de
Sitter spacetime. For vanishing the extra component $X_{\mu\nu}$
or  the Weyl tensor $E_{\mu\nu}$, we have the Minkowski spacetime
on the brane.

Applying the above model, Eq.~(\ref{eq:X-constraint}) becomes
\begin{equation}
    \dot{X}^0_{\ 0} + \left(4 + \beta_{0\lambda}\right)X^0_{\ 0}H
    = \beta_{1\lambda}\frac{k_a}{a^2}H +\beta_3\kappa^2V_{eff}H,
    \label{eq:X-constraint-4}
\end{equation}
where
\begin{eqnarray}
    &&\beta_{0\lambda}=\beta_0 +
    \frac{(1+\gamma)^2}{3+n\gamma}\lambda^2,\quad \beta_{1\lambda}=\beta_1 - \frac{6(1+\gamma)^2}{3+n\gamma}\lambda^2,\nonumber\\
    &&\beta_{2\lambda}=\beta_2 - \frac{(n-1)(3+5n)(1+\gamma)^2}{4(3+n\gamma)}\lambda^2,\nonumber\\
    &&\beta_3=\frac{6n(1-\gamma)^2}{(2+n)(3+n\gamma)}\nonumber\\
    &&~~~~~+\frac{2(1+\gamma)(3+3n+n^2+4n\gamma+9\gamma)}{(2+n)(3+n)(3+n\gamma)}\lambda^2.
\end{eqnarray}
The solution of Eq.~(\ref{eq:X-constraint-4}) is given by
\begin{eqnarray}
    X^0_{\ 0} &=& \frac{\beta_{1\lambda}}{2+\beta_{0\lambda}}\frac{k_a}{a^2}
    +\frac{(3+n)\beta_3}{(3+n)(4 + \beta_{0\lambda})-2\lambda^2}\kappa^2V_{eff}\nonumber\\
    &&  -\frac{\varepsilon_{0\lambda}}{a^{(4 + \beta_{0\lambda})}},
    \label{eq:X-constraint-4-sol}
\end{eqnarray}
where $\varepsilon_{0\lambda}$ is an integration constant. The
Friedmann equation is given by
\begin{eqnarray}
   &&3(1+\alpha_{0\lambda})H^2 =  -3\left[1+ \frac{\beta_{1\lambda}}{3(2+\beta_{0\lambda})}\right]\frac{k_a}{a^2}\nonumber\\
    &&~~~~~~~~+\left[1-\frac{(3+n)\beta_3}{(3+n)(4 +
    \beta_{0\lambda})-2\lambda^2}\right]\frac{\kappa^2V_{eff,0}}{a^{2\lambda^2/(3+n)}}\nonumber\\
    &&~~~~~~~~ +\frac{\varepsilon_{0\lambda}}{a^{(4 + \beta_{0\lambda})}},
    \label{eq:friedmann-vac-2}
\end{eqnarray}
where
\begin{equation}
    \alpha_{0\lambda}=\alpha_{0}-{1\over 6}(1+\gamma)^2\lambda^2.
\end{equation}
We also find that a possible energy leak out of the brane into the
bulk is in the form
\begin{eqnarray}
    &&J_n = -\frac{3(1+\alpha_{0\lambda})}{3+n\gamma}H\dot{\phi}-\frac{(1+\gamma)\lambda}{(3+n\gamma)\kappa}\nonumber\\
    &&\times \left[\frac{n(1+n)+\gamma(12+3n-n^2)}{(2+n)(3+n)(1+\gamma)}\frac{\kappa^2V_{eff,0}}{a^{2\lambda^2/(3+n)}}
    -3\frac{k_a}{a^2}\right].
    \label{eq:leak-1}
\end{eqnarray}
Note that equations (\ref{eq:friedmann-vac-2}) and
(\ref{eq:leak-1}) include five-dimensional case, corresponding to
$n = 0$~\cite{Langlois:2003dd}. By contrast, in higher dimensions,
$J_n$ depends on the effective potential or the induced
cosmological constant on the brane. For example, in the static
internal dimension, $\gamma=0$, and assuming $k_a=0$ for
simplicity, we find
\begin{equation}
    J_n = -\left(1-{\lambda^2\over 6}\right)H_a\dot{\phi}-\frac{n(1+n)\lambda}{3(2+n)(3+n)\kappa}\Lambda_b,
    \label{eq:leak-2}
\end{equation}
where we have used $\kappa^2 V_{eff}=\Lambda_b$. If we impose the
generalized Randall-Sundrum condition, $\Lambda_b=0$, the energy
is flowing onto the brane when $\lambda^2<6$, and flowing out to
the brane when $\lambda^2>6$. For $\Lambda_b>0$, the energy is
flowing onto the brane when $0<\lambda<\sqrt{6}$, and flowing out
to the brane when $\lambda<-\sqrt{6}$.  While for $\Lambda_b<0$,
the energy is flowing onto the brane when $-\sqrt{6}<\lambda<0$,
and flowing out to the brane when $\lambda>\sqrt{6}$.
%%%%%%%%%%%%%%%%%%%%%%%%%%%%%%%%%%%%%%%%%%%%%%%%%%%%%%%%%%%%%%%%%%
%%%%%%%%%%%%%%%%%%%%%% SECTION III %%%%%%%%%%%%%%%%%%%%%%%%%%%%%%%
\section{\label{sec:staticint}Static internal dimension solutions with matter on the
brane}

In the previous section we have focussed upon dilaton-vacuum
solutions on the brane. For completeness we now study the presence
of matter on the brane. In this case we do not have $X^\mu_{\
\mu}=0$. Indeed, we have ${\cal Z}^\mu_{\ \mu}=0$ which it implies
\begin{equation}
    {\cal Z}^0_{\ 0} + 3{\cal Z}^1_{\ 1} + n{\cal Z}^4_{\ 4}=0.
    \label{eq:Z-cons-2}
\end{equation}
Using Eq.~(\ref{eq:x-matter}) and by combining
Eqs.~(\ref{eq:X-cons-1}) and (\ref{eq:Z-cons-2}) we find
\begin{eqnarray}
     &&-n{\cal Z}^4_{\ 4}= {\cal Z}^0_{\ 0} + 3{\cal Z}^1_{\ 1} = -{n\over 2+n}\left(G^0_{\ 0}  + 3G^1_{\ 1} - 2G^4_{\ 4}\right)\nonumber\\
     &&~~~~~ - {2n\over 2+n}\kappa^2V_{eff} + {n\over 2+n}\left(Y^0_{\ 0} + 3Y^1_{\ 1} - 2Y^4_{\ 4}\right),
    \label{eq:Z-cons-3}
\end{eqnarray}
where $Y^{\mu}_{\ \nu}$ is given by Eq.~(\ref{eq:y-matter}) with
the components of $\pi^{\mu}_{\ \nu}$ are
\begin{widetext}
\begin{eqnarray}
    \pi^0_{\ 0} &=& -{1\over 8(3+n)}\left\{2\left(t^0_{\ 0}\right)^2
    +n\left[\left(t^0_{\ 0}\right)^2 - 3 \left(t^1_{\ 1} - t^4_{\ 4}\right)^2\right]\right\}   \ ,
    \label{eq:pi-00}\\
    \pi^i_{\ j} &=& {1\over 8(3+n)}\left\{2\left(t^0_{\ 0}\right)^2- 4t^0_{\ 0}t^1_{\ 1}
    +n\left[\left(t^0_{\ 0}\right)^2  - 2 t^0_{\ 0}t^4_{\ 4}
    + \left(t^1_{\ 1} - t^4_{\ 4}\right)\left(t^1_{\ 1} - 3t^4_{\ 4}\right)\right]\right\} \delta^i_{\ j}\ ,
   \label{eq:pi-ij}\\
    \pi^\alpha_{\ \beta} &=& {1\over 8(3+n)}\left\{2\left(t^0_{\ 0}\right)^2
    - 6t^0_{\ 0}t^1_{\ 1}+ 2t^4_{\ 4}\left(t^0_{\ 0} +3t^1_{\ 1} - 3 t^4_{\ 4}\right)
    +n\left[\left(t^0_{\ 0}\right)^2  - 2t^0_{\ 0}t^4_{\ 4}
     + 3\left(t^1_{\ 1} - t^4_{\ 4}\right)^2\right] \right\} \delta^\alpha_{\ \beta}\ .
   \label{eq:pi-ab}
\end{eqnarray}
\end{widetext}
Here $-t^0_{\ 0}$ is the total energy density, $t^1_{\ 1}=t^2_{\
2}=t^3_{\ 3}$ the total external pressure and $t^4_{\ 4}=t^5_{\
5}=\ldots=t^{3+n}_{\ 3+n}$ the total internal pressure. Inserting
Eq.~(\ref{eq:x-matter}) into Eq.~(\ref{eq:X-constraint}) and then
eliminating ${\cal Z}^1_{\ 1}$ and ${\cal Z}^4_{\ 4}$ by using
Eq.~(\ref{eq:Z-cons-3}), we find
\begin{eqnarray}
    &&\dot{{\cal Z}}^0_{\ 0} + \left(4H_a + nH_b\right) {\cal Z}^0_{\ 0}
    =\kappa^2 J_n \dot{\phi} \nonumber\\
    &&- {n\over 2+n}\left(G^0_{\ 0} + 3G^1_{\ 1} - 2G^4_{\ 4} + 2\kappa^2V_{eff}\right)(H_a-H_b)\nonumber\\
    &&+\dot{Y}^0_{\ 0}+{2\over 2+n}\left[(3+2n)Y^0_{\ 0}-3Y^1_{\ 1}-nY^4_{\ 4}\right]H_a\nonumber\\
    &&+{n\over 2+n}\left[(1+n)Y^0_{\ 0}-3Y^1_{\ 1}-nY^4_{\ 4}\right]H_b.
    \label{eq:Z-constraint-1}
\end{eqnarray}
Note that in the absence of matter on the brane, ${\cal Z}^\mu_{\
\nu}=-X^\mu_{\ \nu}$, Eq.~(\ref{eq:Z-constraint-1}) reduced to
Eq.~(\ref{eq:X-constraint}).

In the following we study the static internal dimension case for
simplicity. In this case we have
\begin{equation}
    H_b = 0, \qquad \text{and} \qquad G^0_{\ 0} + 3G^1_{\ 1} - 2G^4_{\ 4} =0.
    \label{eq:static-cond}
\end{equation}
Then, Eq.~(\ref{eq:Z-constraint-1}) becomes
\begin{eqnarray}
    &&\dot{{\cal Z}}^0_{\ 0} + 4 {\cal Z}^0_{\ 0}H_a
    =\kappa^2 J_n \dot{\phi} - {2n\over 2+n}\kappa^2V_{eff}H_a+\dot{Y}^0_{\ 0}\nonumber\\
    &&~~~~~+{2\over 2+n}\left[(3+2n)Y^0_{\ 0}-3Y^1_{\ 1}-nY^4_{\ 4}\right]H_a.
    \label{eq:Z-constraint-2}
\end{eqnarray}
The conservation of the energy-momentum tensor for the matter
field on the brane and the equation of motion of the scalar field,
respectively, are given by
\begin{eqnarray}
    &&\dot{t}^0_{\ 0} =- 3H_a \left(t^0_{\ 0} -t^1_{\ 1} \right),
    \label{eq:conservation-stat}\\
    && \ddot{\phi} + 3H_a\dot{\phi} + V'_{eff} = -J_n.
    \label{eq:cos-eos-stat}
\end{eqnarray}
Eliminating $Y^\mu_{\ \nu}$ in Eq.~(\ref{eq:Z-constraint-2}), we
obtain
\begin{eqnarray}
    &&\dot{{\cal Z}}^0_{\ 0} + 4 {\cal Z}^0_{\ 0}H_a
    =\kappa^2 J_n \dot{\phi} - {2n\over 2+n}\kappa^2V_{eff}H_a \nonumber\\
    &&+{(2+n)\kappa^4\over 4(3+n)}\sigma'\dot{\phi}t^0_{\ 0}
    + {n\kappa^4\over 4(3+n)}\sigma\left[t^0_{\ 0}+3t^1_{\ 1}-2t^4_{\ 4}\right]H_a\nonumber\\
    &&+\kappa^4\dot{\pi}^0_{\ 0}+{2\over 2+n}\left[(3+2n)\pi^0_{\ 0}-3\pi^1_{\ 1}-n\pi^4_{\ 4}\right]H_a.
    \label{eq:Z-constraint-3}
\end{eqnarray}

We now consider the exponential potential as discussed in the
previous section with static internal dimension. By defining the
equations of state $t^1_{\ 1}=-w_it^0_{\ 0}$, $t^4_{\ 4}=-w_\alpha
t^0_{\ 0}$, and inserting Eq.~(\ref{eq:cos-eos-stat}) into
Eq.~(\ref{eq:Z-constraint-3}) for $J_n$, one can integrate the
following equation,
\begin{widetext}
\begin{eqnarray}
    &&\dot{{\cal Z}}^0_{\ 0} + \left(4 +  {\lambda^2\over 3} \right){\cal Z}^0_{\ 0}H_a
    = -\frac{6n(3+n)+2(3+3n+n^2)b^2}{3(2+n)(3+n)}\kappa^2V_{eff}H_a + 2\lambda^2 {k_a\over a^2}H_a\nonumber\\
    &&~~~~~+\frac{3n(3+n)(1-3w_i+2w_\alpha)+[3+3n+n^2-(3+n)(3w_i+nw_\alpha)]\lambda^2}{12(3+n)^2}\kappa^4\sigma t^0_{\ 0} H_a\nonumber\\
    &&~~~~~+\frac{3n(1-3w_i+3w_\alpha)(w_\alpha+3(1+w_i)(w_i-w_\alpha))+[3w_i+nw_\alpha+3n(w_i-w_\alpha)^2]\lambda^2}{12(3+n)}\kappa^4(t^0_{\ 0})^2 H_a,
    \label{eq:Z-constraint-4}
\end{eqnarray}
\end{widetext}
and yields
\begin{widetext}
\begin{eqnarray}
    &&{\cal Z}^0_{\ 0}=-\frac{6n(3+n)+2(3+3n+n^2)\lambda^2}{(2+n)[12(3+n)+(n-3)\lambda^2]}\kappa^2V_{eff}\nonumber\\
    &&-\frac{3n(3+n)(1-3w_i+2w_\alpha)+[3+3n+n^2-(3+n)(3w_i+nw_\alpha)]\lambda^2}{4(3+n)[3(3+n)(1-3w_i)+n\lambda^2]}\kappa^4\sigma t^0_{\ 0}\nonumber\\
    &&+\frac{3n(1-3w_i+3w_\alpha)[w_\alpha+3(1+w_i)(w_i-w_\alpha)]+[3w_i+nw_\alpha+3n(w_i-w_\alpha)^2]\lambda^2}
    {4(3+n)[6(1+3w_i)-\lambda^2]}\kappa^4(t^0_{\ 0})^2.
    \label{eq:Z-constraint-4-sol}
\end{eqnarray}
\end{widetext}
Inserting Eq.~(\ref{eq:Z-constraint-4-sol}) into (00)-component of
the Einstein equations we find
\begin{widetext}
\begin{eqnarray}
    &&H^2_a + {\bar{k}_a\over a^2}=\beta_4\kappa^2V_{eff}+8\pi G_{eff}\rho+\beta_5\rho^2+{\bar{C}\over a^{4+\lambda^2/3}},
    \label{eq:friedmann-matter}\\
    &&J_n = -\left(1-{\lambda^2\over 6}\right)H_a\dot{\phi}+{\lambda\over \kappa}{k_a\over a^2}
    -\frac{n(1+n)\lambda\kappa}{3(2+n)(3+n)}V_{eff}-\frac{(1-3w_i-nw_\alpha)\lambda\kappa^3}{12(3+n)}\sigma\rho\nonumber\\
    &&~~~~~~+\frac{[4(1+3w_i)+2n(2+2w_i+3w_\alpha+3(w_i-w_\alpha)^2)+n^2(1+2w_i+3(w_i-w_\alpha)^2)]\lambda\kappa^3}{24(2+n)(3+n)}\rho^2,
    \label{eq:leak-stat-matter}
\end{eqnarray}
\end{widetext}
where $\rho=-t^0_{\ 0}$, and
\begin{eqnarray}
    &&\bar{k}_a = \left(1+{\lambda^2\over 6}\right)^{-1}{k}_a,\\
    &&\beta_4 = \frac{2(3+n)(4+n)}{(2+n)[12(3+n)+(n-3)\lambda^2]},\\
    &&8\pi G_{eff}=\frac{(1-3w_i-nw_\alpha)}{2(3+n)(1-3w_i)+ 2n\lambda^2}\kappa^4\sigma,
    \label{eq:g-eff}\\
    &&\beta_5 = \frac{2(1+3w_i) + n(1+2w_\alpha)+
    3(w_i-w_\alpha)^2}{4(3+n)[6(1+3w_i)-\lambda^2]}\kappa^4\\
    &&\bar{C} = {1\over 3}\left(1-{\lambda^2\over 6}\right)^{-1}C.
\end{eqnarray}
The first two terms on the right hand side of
(\ref{eq:friedmann-matter}) are what we would expect for standard
four dimensional cosmology. The third term is quadratic in the
brane energy momentum tensor, and the fourth term is a generalized
dark radiation energy component. Note that in contrast with
five-dimensional case, the effective gravitational constant
depends on time through the brane tension and the equation of
state. One can obtain the variation rate of the effective
gravitational constant is
\begin{equation}
    {\dot{G}_{eff}\over G_{eff}} = -{\lambda^2\over 3+n}H_a.
    \label{eq:g-vary}
\end{equation}
We see that the variation rate of the effective gravitational
constant is smaller than five-dimensional case ($n=0$) with the
exponential potential~\cite{Langlois:2003dd}, and it is relatively
small if the number of the internal dimension $n$ is large.
Observational bounds on ${\dot{G}_{eff}/G_{eff}}$ then constrain
the parameters of the theory. If we assume
$|{\dot{G}_{eff}/(G_{eff}}H_{a,0})|\lesssim 10^{-2}$, where
$H_{a,0}$ is the value of the Hubble parameter today, we have
$\lambda^2 \lesssim (3+n)/100$.

%%%%%%%%%%%%%%%%%%%%%%%%%%%%%%%%%%%%%%%%%%%%%%%%%%%%%%%%%%%%%%%%%%
%%%%%%%%%%%%%%%%%%%%%% SECTION V %%%%%%%%%%%%%%%%%%%%%%%%%%%%%%%%%
\section{\label{sec:E-frame}Cosmological Evolution of the Sources in the Einstein Frame}
In the previous sections, we have discussed the cosmological
evolutions in the induced metric frame, which is given by Eq.~(\ref{eq:einstein-brane}).
In this section, we study the 4-dimensional cosmological evolution in the Einstein frame and discuss the properties of the resulting
4-dimensional cosmological evolution in this frame. A complementary way of studying higher-dimensional theories is to dimensionally reduce the action by integrating out the internal dimensions and then perform a conformal transformation.

From Eq.~(\ref{eq:einstein-brane}) we define the Newton gravitational constant in $(4+n)$-dimensions as
\begin{equation}
    8\pi G_{4+n} = \frac{(2+n)\kappa^4}{4(3+n)}\sigma.
    \label{eq:4+n-dnewton}
\end{equation}
If the brane tension is not depending on the bulk scalar field (minimally coupled),
the $(4+n)$-dimensional Newton gravitational constant is truly constant. In this section we assume $\sigma=const.$
Note that recovering the 4-dimensional effective Newton gravitational constant requires
an additional time-dependent factor,
\begin{equation}
    8\pi G_{4} = \frac{8\pi G_{4+n}}{V_n},
    \label{eq:4-dnewton}
\end{equation}
where $V_n$ denotes the volume of the $n$-dimensional internal
spaces, while the 4-dimensional cosmological constant is given by
\begin{equation}
    \Lambda_4 = V_n \Lambda_b,
    \label{eq:4-dcc}
\end{equation}
where $\Lambda_b$ is given by Eq.~(\ref{eq:cos-brane}). Note that both $G_4$ and $\Lambda_4$ are time-dependent. However, it must be remembered that we are not in the Einstein frame. Similarly, the 4-dimensional energy density is given by $\rho_4=V_n\rho$.

One can compare the time-dependent of the effective Newton gravitational constant Eqs.~(\ref{eq:g-eff}) and (\ref{eq:4-dnewton}). We find
\begin{equation}
    G_{4}=\frac{(2+n)[2(3+n)(1-3w_i)+ 2n\lambda^2]}{4(3+n)(1-3w_i-nw_\alpha)}{G_{eff}\over V_n}.
\end{equation}
Here, $V_n$ to be a constant, which corresponding to the static internal dimensions. As we expected 
\begin{equation}
    {\dot{G}_{4}\over G_{4}}={\dot{G}_{eff}\over G_{eff}}.
    \label{eq:g-vary-1}
\end{equation}
However, for the static internal dimensions case with a constant brane tension, the effective Newton gravitational constant is exactly constant. It depends on the equations of state, the potential parameter $\lambda$ and the number of internal dimensions.

In order to recovering 4-dimensional cosmological evolution in the Einstein frame, we proceed through three steps: (1) a dimensional reduction of the $(4+n)$-action by integrating out the internal dimensions; (2) a conformal transformation of the 4-dimensional induced metric into Einstein frame; (3) a redefinition of the scale factor $b$ into a new scalar field to give its kinetic term a canonical form. From the gravitational sector, the resulting 4-dimensional Einstein frame action is given by
\begin{eqnarray}
   S_{*} &=& \int d^4 x \sqrt{-h_{*}}\left[\frac{1}{16\pi G_{4*}}
   R_{*}-{1\over 2}h^{*\mu\nu}\partial_\mu\Phi\partial_\nu\Phi\right.\nonumber\\
   &&\left.+V_{*}(\Phi) \right],
\end{eqnarray}
where we mark the quantities in the Einstein frame with the subscript $*$ and a new scalar field is defined by
\begin{equation}
    \Phi = \sqrt{\frac{n(2+n)}{16\pi G_{4*}}}~\ln b.
\end{equation}
We also have the relations between the variables in the 4-dimensional induced metric frame and those
in the Einstein frame as
\begin{equation}
    dt_{*} = b^{n/2} dt,\quad a_{*}=b^{n/2} a,
\end{equation}
where the scale factor $b$ is appropriately expressed in $\Phi$. Hence we have the relation
\begin{equation}
    H_{a*}  = b^{-n/2} \left(H_a+{n\over 2}H_b\right).
\end{equation}
where $H_{a*}$ is the Hubble parameter in the Einstein frame,
\begin{equation}
    H_{a*} = {1\over a_{*}}\frac{da_{*}}{dt_{*}}.
\end{equation}
In the static internal dimensions case, $b=b_0=1$, we have $a_{*}=a$ so that $H_{a*} = H_{a}$.

From the matter sector, we have
\begin{equation}
    \rho_{*} = b^{-2n}\rho_4.
\end{equation}
Assuming $V_n\sim b^n$, the evolution of the 4-dimensional energy density in the Einstein frame is given by
\begin{equation}
    \rho_{*} \propto b^{-{n\over 2}(1-3w_i+2w_\alpha)}a_{*}^{-3(1+w_i)}.
\end{equation}
Note that we recover standard 4-dimensional energy density for $n=0$ or if the equations of state satisfy the following constraint:
\begin{equation}
    1-3w_i+2w_\alpha=0.
    \label{eq:stat-cond}
\end{equation}
For instance, in the radiation-dominated era $w_i=1/3$, we find $w_\alpha=0$, the internal pressure drops to zero. Moreover, in order to understanding the physical interpretations of the constraint (\ref{eq:stat-cond}), let us consider equation (\ref{eq:static-cond}). Since the components of the Einstein tensor are corresponding to the matter field: $G^0_{\ 0}\sim t^0_{\ 0}$, $G^1_{\ 1}\sim t^1_{\ 1}$, and $G^4_{\ 4}\sim t^4_{\ 4}$, and using the relations $t^1_{\ 1}=-w_i t^0_{\ 0}$, $t^4_{\ 4}=-w_\alpha t^0_{\ 0}$, the constraint (\ref{eq:stat-cond}) can be interpreted as a necessary condition for static internal dimensions.

We have shown that we can obtain the standard cosmology in the Einstein frame by assuming a constant brane tension.
In the case of the non-minimally coupled to bulk scalar field, in which the brane tension is a function of the bulk scalar field, a conformal transformation is determined by the form of brane tension. For instance, if the brane tension is given by Eq.~(\ref{eq:tension}), the Einstein frame is obtained by a conformal transformation of the 4-dimensional metric: $h_{*\mu\nu}=b^n e^{2\varphi}h_{\mu\nu}$, where $2\varphi=\lambda\kappa\phi/(3+n)$. Then the Hubble parameter in the Einstein frame is related to the Hubble parameter in the induced metric
frame as
\begin{equation}
    H_{a*}  = b^{-n/2}e^{-\varphi} \left(H_a+{n\over 2}H_b + \dot{\varphi}\right),
\end{equation}
and the evolution of the 4-dimensional energy density in the Einstein frame is given by $\rho_{*} = b^{-n}e^{-4\varphi}\rho$, with $\rho$ is the higher-dimensional energy density.
%%%%%%%%%%%%%%%%%%%%%%%%%%%%%%%%%%%%%%%%%%%%%%%%%%%%%%%%%%%%%%%%%%
%%%%%%%%%%%%%%%%%%%%%% CONCLUSION %%%%%%%%%%%%%%%%%%%%%%%%%%%%%%%%
\section{\label{sec:conclusion}Conclusions}
In this paper we studied the brane-world cosmological model in
higher-dimensional spacetime with a bulk scalar field. The
$(4+n)$-dimensional gravitational field equations can be
formulated to standard form with the extra component. In the
absence of the matter on the brane, we can interpret the extra
component as the energy-momentum tensor for the dark radiation. We
used this formalism to discuss cosmology in the brane Universe in
the context of the Kaluza-Klein brane-world scheme, i.e., to
consider Kaluza-Klein compactifications on the brane. By assuming
the cosmological symmetry on the Kaluza-Klein brane, we derived
the Friedmann equation and a possible energy leak out of the brane
into the bulk.

If the bulk and brane potentials obey the generalized
Randall-Sundrum condition and the energy conservation for scalar
field on the brane is satisfied, we find that the brane Universe
starts expanding away from the initial singularity both for the
internal scale factor is proportional to the external scale factor
and the static internal scale factor. A significant result in
present study is that the possibility of energy flow out of the
brane into the bulk, which it depends on the effective potential
or the effective cosmological constant on the brane. It is
different with the five-dimensional dilatonic brane-world
cosmologies. By constraining the potential parameter $\lambda$ and
the effective cosmological constant, there exist two possibilities
of the energy flows: onto and out to the brane. Finally, we derive
the Friedmann equation with matter on the brane. We find that the
effective gravitational constant depends on time through the brane
tension and the equation of state. However, after appropriate conformal transformation,
the 4-dimensional Newton gravitational constant in the Einstein frame is true constant.
A new result in Section~\ref{sec:E-frame} is that we recover standard 4-dimensional energy density
in the Einstein frame if the equations of state satisfy the constraint Eq.~(\ref{eq:stat-cond}).

There would be various extensions of our considerations. By
introducing a bulk scalar field, it is interesting to stabilize
the distance between the two Kaluza-Klein branes as is done in the
context of the first model introduced by Randall and Sundrum. It
also important to understand inflation without inflaton and the
quantum fluctuations of brane-inflation in the Kaluza-Klein brane
scenario with a bulk scalar field. It is intriguing to consider
the low energy description of this brane-world model
\cite{Feranie:2010th}. We leave these issues for future studies.

%===============================================================%
%************************ ACKNOWLEDGEMENT **********************%
%===============================================================%
\section{Acknowledgement}
This work was supported by Hibah Kompetensi DIKTI 2011 No.
2511/H14/HM/2011.
%===============================================================%
%************************ REFERENCES ***************************%
%===============================================================%

\end{document}